\documentclass[twocolumn,nofootinbib,amsmath,amssymb,a4paper]{revtex4}

\usepackage{graphicx}
\usepackage{dcolumn}
\usepackage{bm}

\begin{document}

\title{Rare K decays in the Standard Model}
\author{Christopher Smith\footnote{Electronic address: chsmith@itp.unibe.ch}}
\affiliation{Institut f\"{u}r Theoretische Physik, Universit\"{a}t Bern, CH-3012
Bern, Switzerland}

\begin{abstract}
The very clean theoretical predictions for the rare decays
$K\rightarrow\pi\nu\bar{\nu}$ and $K_{L}\rightarrow\pi^{0}\ell^{+}\ell^{-}$
are reviewed, and their various theoretical inputs summarized. The less 
favorable situation for $K_{L}\rightarrow\mu^{+}\mu^{-}$ is also commented.
\end{abstract}

\maketitle

\section{Introduction}

The rare decays $K\rightarrow\pi\nu\bar{\nu}$ and $K_{L}\rightarrow\pi^{0}\ell
^{+}\ell^{-}$, driven by semi-leptonic flavor-changing neutral currents (FCNC),
are exceptionally clean probes of the flavor structure of the Standard Model, 
or of the still elusive New Physics. Concerted theoretical efforts have brought 
the SM predictions to an impressive level of accuracy (for the current experimental situation, see \cite{exp}). 
In this section, the main theoretical ingredients are briefly reviewed, while the situation for each
mode is summarized in the following sections.

\subsection{FCNC electroweak structure}

FCNC arise at one loop in the electroweak theory. The processes driving
the rare semi-leptonic $K$ decays are the $W$ box, $Z$ and $\gamma$
penguins\cite{BuchallaBL96}, see Fig.1, and lead to the amplitudes%
\begin{align*}
\mathcal{A}\left(  K_{L}\rightarrow\pi^{0}X\right)    & =\sum_{q=u,c,t}\left(
\operatorname{Im}\lambda_{q}+\varepsilon\operatorname{Re}\lambda_{q}\right)
y_{q}^{X}\left(  m_{q}\right)  \,,\\
\mathcal{A}\left(  K^{+}\rightarrow\pi^{+}X\right)    & =\sum_{q=u,c,t}\left(
\operatorname{Re}\lambda_{q}+i\operatorname{Im}\lambda_{q}\right)  y_{q}%
^{X}\left(  m_{q}\right)  \,,
\end{align*}%
with $X=\nu\bar{\nu},\ell^{+}\ell^{-}$ and $\lambda_{q}=V_{qs}^{\ast}V_{qd}$. In
standard terminology, the $\varepsilon$ part is the indirect CP-violating
piece (ICPV), while the $\operatorname{Im}\lambda_{q}$ part is called direct
CP-violating (DCPV). The amplitude for $K_{S}$ is obtained from the $K_{L}$
one by interchanging $\operatorname{Im}\lambda_{q}\leftrightarrow
\operatorname{Re}\lambda_{q}$.

Without the dependence of the loop functions $y_{q}^{X}$ on the quark masses,
CKM unitarity would imply vanishing FCNC (GIM mechanism). Now, looking at
these dependences, combined with the scaling of the CKM elements, one can
readily get a handle on the importance of each quark contribution, and thereby
on the cleanness achievable for the decay once QCD effects are included.%

\begin{figure*}[t]
\includegraphics[width=0.88\textwidth]{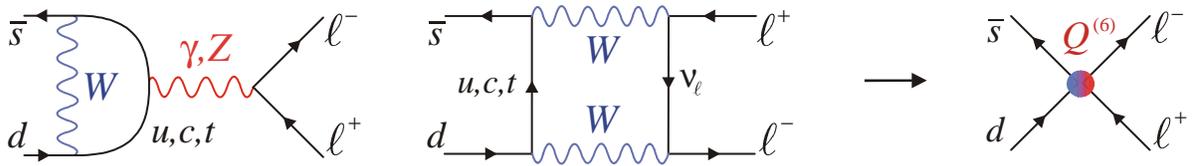}
\caption{The $Z,\gamma$ penguin and $W$ box generating the effective FCNC interactions
relevant for rare $K$ decays.}
\end{figure*}

For $X=\nu\bar{\nu}$, only the $Z$ penguin and $W$ box enter, $y_{q}^{\nu
\bar{\nu}}\sim m_{q}^{2}$, and light-quark contributions are suppressed.
Since, in addition, $\varepsilon\sim10^{-3}$ and $\operatorname{Re}\lambda
_{t}\sim\operatorname{Im}\lambda_{t}$, ICPV is very small. For $K^{+}$, the
$c$-quark contribution is suppressed from the loop, but enhanced by
$\operatorname{Re}\lambda_{c}>>\lambda_{t}$, and ends up being comparable to
the $t$-quark contribution.

For $X=\ell^{+}\ell^{-}$, the photon penguin also enters with its scaling
$y_{q}^{\ell\ell}\sim\log(m_{q})$ for $m_{q}\rightarrow0$. In the standard CKM
phase-convention, DCPV is still short-distance dominated thanks to
$\operatorname{Im}\lambda_{u}=0$, but not ICPV, completely dominated by the
long-distance $u$-quark photon penguin, $K_{1}\rightarrow\pi^{0}\gamma^{\ast
}\rightarrow\pi^{0}\ell^{+}\ell^{-}$. The same holds for $K^{+}%
\rightarrow\pi^{+}\ell^{+}\ell^{-}$, completely dominated by long-distance and
therefore not very interesting for New Physics search.

For $K_{L}\rightarrow\ell^{+}\ell^{-}$, there is no photon penguin, and the
electroweak structure is similar to $K\rightarrow\pi\nu\bar{\nu}$, up to the change
$\operatorname{Im}\lambda_{q}\leftrightarrow\operatorname{Re}\lambda_{q}$.

Along with these contributions, there can be two-loop, third order
electroweak contributions, if the extra suppression is compensated by
non-perturbative long-distance enhancement. This occurs for modes with charged
leptons, where the double-photon penguin gives a CP-conserving contribution
($\sim\operatorname{Re}\lambda_{q}$) to $K_{L}\rightarrow\pi^{0}\ell^{+}%
\ell^{-}$ and $K_{L}\rightarrow\ell^{+}\ell^{-}$, and is completely dominated
by long-distance ($u$-quark), Fig.2c.

\subsection{QCD corrections}

Having identified the relevant electroweak structures, QCD effects have now to
be included. This is done in three main steps:%

\begin{figure*}
\includegraphics[width=0.95\textwidth]{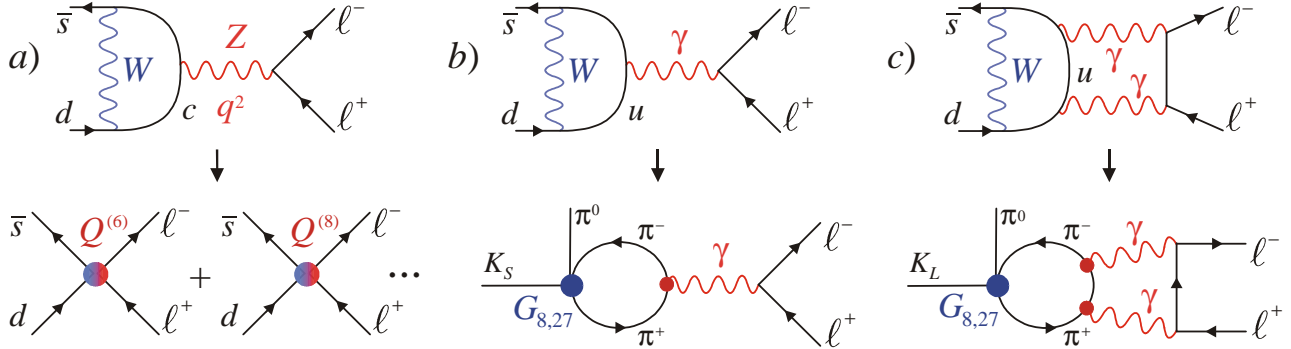}
\caption
{Illustration of the treatment of various QCD effects. a) Tower of effective FCNC
interactions generated by the $c$-quark integration. b) Non-local long-distance
photon penguin with $u$-quark contribution represented by meson loops.
c) Idem as b) for the double-photon penguin.}
\end{figure*}

\textit{Step 1}: Integration of heavy degrees of freedom (top, $W$, $Z$),
including perturbative QCD effects above $M_{W}$. This generates local FCNC
operators (Fig.1 with $t$-quark), and Fermi-type four-fermion local operators.

\textit{Step 2}: Resummation of QCD corrections (running down). At the $c$
threshold (similar for $b,\tau$), four-fermion operators are combined to form
closed $c$-loops, which are then replaced by a tower of effective interactions
in increasing powers of (external momentum)/(charm mass), Fig.2a. The lowest
order consists again of the dimension-six FCNC operators, while dimension-eight
operators are corrections scaling naively like $m_{K}^{2}/m_{c}^{2}\sim15\%$.

These first two steps (the OPE) can, in principle, be achieved to any desired
level of precision within perturbative QCD, though the computation of the
required multiloop diagrams represents a formidable task at higher orders.
Still, this is unavoidable in order to reduce theoretical errors, in
particular scale dependences. At this stage, one has obtained the complete
Hamiltonian, i.e. all the effective operators, with the short-distance physics
encoded in their Wilson coefficients.

\textit{Step 3}: To get the amplitudes, the matrix elements of these 
operators between meson states remain to be estimated. To this end, one 
makes use of the symmetries of QCD, as embodied in Chiral Perturbation 
Theory (ChPT), to relate the desired matrix elements to experimentally 
known quantities.

For the most interesting dimension-six semi-leptonic operators, the matrix
elements can be related to those of $K_{\ell2},K_{\ell3}$ decays
(taking into account isospin-breaking corrections). Contributions from four 
light-quark operators ($Q_{1},...,Q_{6}$) are represented directly in terms of meson
fields in ChPT, such that non-local $u$-quark loops are represented as meson
loops (Fig.2b,c). The price to pay is the introduction of some unknown low-energy
constants ($G_{8,27,...}$), to be extracted from experiment. In particular,
$G_{8}$ is fixed from $\mathcal{B}\left( K\rightarrow\pi\pi\right)$, accounting 
for the large non-perturbative $\Delta I=1/2$ effects. For dimension-eight 
operators, an approximate matching is done with the ChPT representation of the
$u$-quark contributions.

\section{The $K^+\rightarrow\pi^+\nu\bar{\nu}$ and $K_L\rightarrow\pi^0\nu\bar{\nu}$ decays}

Thanks to the suppression of light-quark effects, these modes are the cleanest 
and their rates are precisely predicted within the SM. Their branching ratios read:
\begin{gather*}
\mathcal{B}_{+}^{\nu\bar{\nu}}=\kappa_{+}\left(  \left|
\frac{\operatorname{Im}\lambda_{t}}{\lambda^{5}}X_{t}\right|  ^{2}+\left|
\frac{\operatorname{Re}\lambda_{t}}{\lambda^{5}}X_{t}+\frac{\operatorname{Re}%
\lambda_{c}}{\lambda}P_{u,c}\right|  ^{2}\right)  \,,\\
\mathcal{B}_{L}^{\nu\bar{\nu}}=\kappa_{L}\left|  \frac{\operatorname{Im}%
\lambda_{t}}{\lambda^{5}}X_{t}\right|  ^{2}\,,
\end{gather*}
with $P_{u,c}=P_{c}+\delta P_{u,c}$. The Wilson coefficient of the dimension-six
FCNC operator $Q^{\nu}=\left(  \bar{s}d\right)  _{V-A}\left(  \bar{\nu}%
\nu\right)  _{V-A}$ arising from the top-quark loop is known at NLO, $X_{t}%
=1.646\pm0.041$\cite{BurasGHN05}. For the charm-quark, the contribution to this 
operator has recently been obtained at NNLO\cite{BurasGHN05}, $P_{c}=0.37\pm0.04$. 
Residual $c$-quark effects from dimension-eight operators\cite{FalkLP01}, along
with long-distance $u$-quark effects\cite{LDKpinn} amount to a small correction
$\delta P_{u,c}=0.04\pm0.02$\cite{IsidoriMS05}. The matrix elements of $Q^{\nu}$ 
are known from $K_{\ell3}$, including the leading isospin-breaking corrections
\cite{MarcianoP96}, and are encoded into $\kappa_{L}=2.29\pm0.03\cdot10^{-10}$
and $\kappa_{+}=5.26\pm0.06\cdot10^{-11}$ for $\lambda=0.225$. Finally, for 
$K_{L}\rightarrow\pi^{0}\nu\bar{\nu}$, ICPV is of about 1\%\cite{BuchallaB96}
while the CP-conserving contribution arising from box diagrams is less 
than 0.01\%\cite{BuchallaI98}.%

\begin{table*}[tb]
\caption{Coefficients encoding the various contributions to $\mathcal
{B}^{\ell^+\ell^-}$}
\label{tab:wide}\begin{tabular}{ccccc}
\hline& $C_{dir}^{\ell}$ & $C_{int}^{\ell}$ & $C_{mix}^{\ell}$ & $C_{\gamma
\gamma}^{\ell}$ \\ \hline $\ell=e$\hspace{2mm} & $\left(4.62\pm0.24\right) \;\left(
y_{7V}^{2}+y_{7A}^{2}\right) $ & $\left( 11.3\pm0.3\right) \;y_{7V}%
$ & $14.5\pm0.5,$ & $\approx0$  \\
$\ell=\mu$\hspace{2mm} & $\left( 1.09\pm0.05\right) \left(
y_{7V}^{2}+2.32y_{7A}^{2}\right) $ & $\left( 2.63\pm0.06\right) \;y_{7V}$ &
$3.36\pm0.20$ & $5.2\pm1.6$ \\ \hline\end{tabular}\end{table*}

The SM predictions are then
\begin{align*}
\mathcal{B}\left(  K_{L}\rightarrow\pi^{0}\nu\bar{\nu}\right)   &
=(2.7\pm0.4)\cdot10^{-11}\,,\\
\mathcal{B}\left(  K^{+}\rightarrow\pi^{+}\nu\bar{\nu}\right)   &
=(8.4\pm1.0)\cdot10^{-11}\,.
\end{align*}
The error on $K_{L}\rightarrow\pi^{0}\nu\bar{\nu}$ is dominated by
$\operatorname{Im}\lambda_{t}$, while for $K^{+}\rightarrow\pi^{+}\nu\bar{\nu
}$, it breaks down to scales (13\%), $m_{c}$(22\%), CKM,
$\alpha_{S}$, $m_{t}$ (37\%) and matrix-elements from $K_{\ell3}$ and
light-quark contributions (28\%)\cite{BurasGHN05}. Further improvements are thus possible
through a better knowledge of $m_{c}$, of the isospin-breaking in the
$K\rightarrow\pi$ form-factors, or by a lattice study of higher-dimensional
operators\cite{IsidoriMT05}.

As the determination of $\lambda_{t}$ from general UT fits to B physics data
is already very precise, and expected to be further improved in the near future, 
the main interest of the $K\rightarrow\pi\nu\bar{\nu}$ decays
is to test the CKM paradigm for CP-violation in the SM. Indeed, these modes do offer
a particularly interesting independent determination since, as discussed in \cite{Mescia07}, 
they are very sensitive to a large class of New Physics models. As such, they 
also constitute one of the best windows into the flavor structure of the New 
Physics that will hopefully be uncovered at LHC.

\section{The $K_{L}\rightarrow\pi^{0}\ell^{+}\ell^{-}$ decays}

Here the situation is more involved. The $t$ and $c$-quark contributions
generate both the dimension-six vector $Q_{7V}=(\bar{s}d)_{V}(\bar{\ell}%
\ell)_{V}$ and axial-vector $Q_{7A}=(\bar{s}d)_{V}(\bar{\ell}\ell)_{A}$
operators, whose Wilson coefficients $y_{7V,7A}$ are known to
NLO\cite{BuchallaBL96}. The former produces the $\ell^{+}\ell^{-}$ pair in a
$1^{--}$ state, the latter in a $1^{++}$ state and, in addition, for $\ell=\mu
$, in a helicity-suppressed $0^{-+}$ state.

Indirect CP-violation is related to $K_{S}\rightarrow\pi^{0}\ell^{+}\ell^{-}$,
for which the long-distance photon penguin dominates (Fig.2b). In ChPT, loops
are small and one just needs to fix a counterterm, $a_{S}$\cite{DambrosioEIP98}.
This can be done up to a sign from NA48 measurements as $\left|
a_{S}\right|  =1.2\pm0.2$\cite{NA48_KSpill}. Producing $\ell^{+}\ell^{-}$ in a
$1^{--}$ state, it interferes with the contribution from $Q_{7V}$, arguably
constructively\cite{BuchallaDI03,FriotGD04}.%

\begin{figure*}
\includegraphics[width=0.98\textwidth]{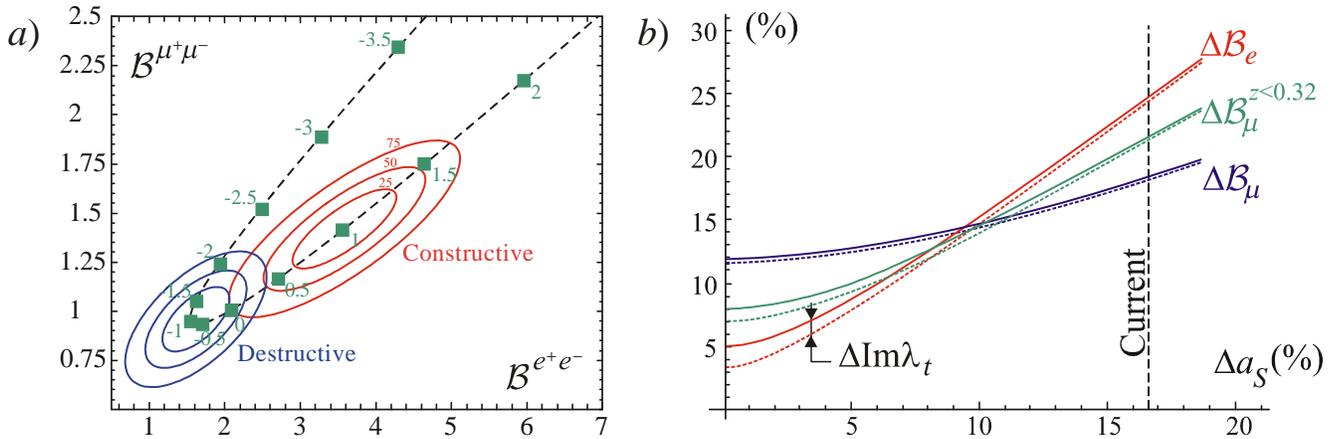}
\caption{a) $\mathcal{B}^{\mu^+\mu^-}$ against $\mathcal{B}^{e^+e^-}%
$, in units of $10^{-11}$. The hyperbola denotes common rescalings of $y_{7A,7V}$ 
(or $\operatorname{Im}\lambda_t$), while the 25, 50 and 75\% confidence regions 
correspond to the current SM predictions for constructive and destructive ICPV-DCPV 
interference. b) Evolution of the error on $\mathcal{B}^{\mu^+\mu^-}$ and $\mathcal
{B}^{e^+e^-}$ as a function of  the error on $a_S$. The residual error due to
$\operatorname{Im}\lambda_t$ is smaller. The middle curve indicates the 
improvement achievable by selecting events with muon invariant-mass smaller than 
$2m_\pi $, which amounts to cutting away the bulk of the two-photon CPC contribution.}
\end{figure*}

The CP-conserving (CPC) contribution from $Q_{1,...,6}$ proceeds through
two-photons, i.e. produces the lepton pair in either a helicity-suppressed
$0^{++}$ or phase-space suppressed $2^{++}$ state. Only the $0^{++}$ state is
produced at LO through the finite two-loop process $K_{L}\rightarrow\pi
^{0}P^{+}P^{-}\rightarrow\pi^{0}\gamma\gamma\rightarrow\pi^{0}\ell^{+}\ell
^{-}$, $P=\pi,K$ (Fig.2c). Higher order corrections are estimated using
$K_{L}\rightarrow\pi^{0}\gamma\gamma$ experimental data for both the $0^{++}%
$\cite{IsidoriSU04} and $2^{++}$ contributions\cite{BuchallaDI03}.

Altogether, the branching ratios are
\[
\mathcal{B}^{\ell^{+}\ell^{-}}=(C_{dir}^{\ell}\pm C_{int}^{\ell}\left|
a_{S}\right|  +C_{mix}^{\ell}\left|  a_{S}\right|  ^{2}+C_{\gamma\gamma}%
^{\ell})\cdot10^{-12}\,,
\]
with the coefficients given in Table I. Interestingly, these coefficients obey
$C_{i}^{\mu}/C_{i}^{e}\approx0.23$ due to the phase-space suppression, but
for the helicity-suppressed contributions arising from the $Q_{7A}$ operator
(DCPV) and from $\gamma\gamma$ (CPC). This maintains the sensitivity of
$\mathcal{B}^{\mu^{+}\mu^{-}}$ on the interesting short-distance physics at
the same level as $\mathcal{B}^{e^{+}e^{-}}$. Further, it allows in principle
to disentangle the $Q_{7V}$ and $Q_{7A}$ contributions from the measurements
of both modes. This is illustrated in Fig.3a, where the hyperbola
corresponds to a common rescaling of both $y_{7A}$ and $y_{7V}$%
\cite{IsidoriSU04}. As discussed in \cite{IsidoriSU04,MesciaST06}, this plane is 
particularly interesting to look for signals of New Physics, and identify 
its precise nature\cite{Mescia07}.

In the SM, $y_{7A}\left(  M_{W}\right)  =-0.68\pm0.03$ and $y_{7V}\left(
\mu\approx1\;\text{GeV}\right)  =0.73\pm0.04$\cite{BuchallaBL96}, and the
predicted rates are\cite{BuchallaDI03,IsidoriSU04,MesciaST06}:%
\begin{align*}
\mathcal{B}\left(  K_{L}\rightarrow\pi^{0}e^{+}e^{-}\right)   & =3.54_{-0.85}%
^{+0.98}\;\left(  1.56_{-0.49}^{+0.62}\right)  \cdot10^{-11}\;,\\
\mathcal{B}\left(  K_{L}\rightarrow\pi^{0}\mu^{+}\mu^{-}\right)   &
=1.41_{-0.26}^{+0.28}\;\left(  0.95_{-0.21}^{+0.22}\right)  \cdot10^{-11}\,,
\end{align*}
for constructive (destructive) interference (Fig.3a). Overall, the error on
$a_{S}$ is currently the most limiting and better measurements of
$K_{S}\rightarrow\pi^{0}\ell^{+}\ell^{-}$ would be welcomed (Fig.3b). Also,
better measurements of $K_{L}\rightarrow\pi^{0}\gamma\gamma$ would help reduce
the error on the $0^{++}$ and $2^{++}$ contributions. Alternatively, they can
be partially cut away through energy cuts or Dalitz plot
analyses\cite{BuchallaDI03,IsidoriSU04,MesciaST06}.

The integrated forward-backward (or lepton-energy) asymmetry
\cite{MesciaST06,AFB}
\[
A_{FB}^{\ell}=\frac{N\left(  E_{\ell^{-}}>E_{\ell^{+}}\right)  -N\left(
E_{\ell^{-}}<E_{\ell^{+}}\right)  }{N\left(  E_{\ell^{-}}>E_{\ell^{+}}\right)
+N\left(  E_{\ell^{-}}<E_{\ell^{+}}\right)  }\,,
\]
is generated by the interference between CP-conserving and CP-violating
amplitudes. It cannot be reliably estimated at present for $\ell=e$ because of
the poor theoretical control on the $2^{++}$ contribution. The situation is
better for $\ell=\mu$, for which this part is negligible,%
\[
A_{FB}^{\mu}=\left(  1.3\left(  1\right)  y_{7V}\pm1.7(2)\left|  a_{S}\right|
\right)  \cdot10^{-12}\,\text{\thinspace}/\,\mathcal{B}^{\mu^{+}\mu^{-}}\,,
\]
i.e., $A_{FB,\mathrm{SM}}^{\mu}\approx20\%$($-12\%$) for constructive
(destructive) interference\cite{MesciaST06}. Interestingly, though the error is large,
$A_{FB}^{\mu}$ can be used to fix the sign of $a_{S}$.

\section{The $K_{L}\rightarrow\mu^{+}\mu^{-}$ decay}

For this mode, the short-distance (SD) piece from $t$ and $c$-quarks is known 
to NLO and NNLO\cite{GorbahnH06}, respectively. Indirect
CP-violation is negligible. The long-distance (LD) contribution from $Q_{1,...,6}$
matrix elements proceeds again through two-photons. Still,
there are three differences with respect to the $K_{L}\rightarrow\pi^{0}\ell
^{+}\ell^{-}$ decays.

First, the contribution from the imaginary part of the $\gamma\gamma$ loop,
estimated from $K_{L}\rightarrow\gamma\gamma$, is much larger than SD, and
already accounts for the bulk of the experimental $K_{L}\rightarrow\mu^{+}%
\mu^{-}$ rate. Second, while the charged meson loop in $K_{L}\rightarrow\pi^{0}\ell^{+}%
\ell^{-}$ acts like a cut-off, and a finite result is found, now the two
photons arise from the axial anomaly, and $K_{L}\rightarrow\pi^{0},\eta
,\eta^{\prime}\rightarrow\gamma\gamma\rightarrow\mu^{+}\mu^{-}$ is divergent, 
requiring unknown counterterms. To estimate them, though still with a large error, 
one can use the experimental information on $K_{L}\rightarrow\gamma^{\ast}\gamma^{\ast}$ 
together with the perturbative behavior of the $\bar{s}d\rightarrow
\bar{u}u\rightarrow\gamma\gamma$ loop\cite{IsidoriU03}. Finally, SD and LD
produce the same $0^{-+}$ state and thus interfere. This interference, which
depends on the sign of $\mathcal{A}(K_{L}\rightarrow\gamma\gamma)$, is
presumably constructive\cite{GerardST05}. Better measurements of
$K_{S}\rightarrow\pi^{0}\gamma\gamma$ or $K^{+}\rightarrow\pi^{+}\gamma\gamma$
could settle this sign.

$K_{L}\rightarrow\mu^{+}\mu^{-}$ is thus obviously not as clean as $K\rightarrow\pi\nu\bar{\nu}$
or $K_{L}\rightarrow\pi^{0}\ell^{+}\ell^{-}$. 
Nevertheless, being measured precisely, it can still lead to interesting 
constraints in some specific scenarios like SUSY at large $\tan\beta$\cite{Mescia07}.

\section{Conclusion}

Thanks to the numerous theoretical efforts, the four rare decays, 
$K\rightarrow\pi\nu\bar{\nu}$ and $K_{L}\rightarrow\pi^{0}\ell^{+}\ell^{-}$,
now  provide for one of the cleanest and most sensitive tests of the Standard Model. 
These modes are promising not only to get clear signals of New Physics 
-- or to severely constrain it --, but also to uncover the nature of the possible 
New Physics at play through the specific pattern of deviations they would exhibit 
with respect to the SM predictions.

\begin{acknowledgments}
I wish to thank the convenors of WG3 for the kind invitation. This work is
supported by the Schweizerischer Nationalfonds.
\end{acknowledgments}

\end{document}